\documentclass[prl,twocolumn,showpacs]{revtex4}

\usepackage{graphicx}
\usepackage{t1enc}
\begin{document}

\title{Phase Field Theory of Heterogeneous Crystal Nucleation}

\author{L\'aszl\'o Gr\'an\'asy,$^1$ Tam\'as Pusztai,$^1$,
David Saylor,$^2$ and James A. Warren$^3$}
\affiliation{$^1$Research Institute for Solid State Physics and Optics,
   H-1525 Budapest, P.O.B.49, Hungary}
\affiliation{$^2$Food and Drug Administration, Rockville, MD 20852}
\affiliation{$^3$National Institute of Standards and Technology,
Gaithersburg, MD 20889}

\date{\today}

\begin{abstract}
  A phase field approach is developed to model wetting and
  heterogeneous crystal nucleation of an undercooled pure liquid in
  contact with a sharp wall. We discuss various choices for the
  boundary condition at the wall and determine the properties of
  critical nuclei, including their free energy of formation and the
  contact angle as a function of undercooling. We find for particular
  choices of boundary conditions, we may realize either an analog of
  the classical spherical cap model or decidedly non-classical behavior,
  where the contact angle decreases from its value taken at the
  melting point towards complete wetting at a critical undercooling.
\end{abstract}


\maketitle

Heterogeneous nucleation is not only a phenomenon of classic
importance in materials science, but also remains one of continuously
growing interest, due to the emerging technological interest in
nanopatterning techniques and control of related nanoscale processes
\cite{ref0}. While solidification of pure undercooled liquids is
initiated by homogeneous nucleation (the formation of small
crystalline fluctuations exceeding a critical size determined by the
interplay of the driving force of crystallization and the interfacial
free energy \cite{ref1}) the presence of foreign particles, container
walls, and other heterogeneities typically facilitates this process
\cite{ref2}. Despite its vast technological importance, heterogeneous
nucleation remains poorly understood. This deficit stems from the
complexity of describing the interaction between the foreign matter
and the solidifying melt.

Wetting of a foreign wall by fluids/crystals has been studied
extensively \cite{ref3} including such phenomena as critical wetting
and phase transitions at interfaces \cite{ref4}.
Various methods have been applied to address these problems such as
continuum models \cite{ref6} and atomistic simulations \cite{ref7}.
Despite this inventory, recent studies \cite{ref8} addressing
heterogeneous crystal nucleation rely almost exclusively on the
classical spherical cap model, which assumes mathematically sharp
interfaces \cite{ref9}. Here the wall-liquid and wall-solid
interactions are characterized by the contact angle $\theta$ that is
determined from the interfacial free energies by Young's equation:
$\gamma_{WL} = \gamma_ {WS} + \gamma_{SL} \cos(\theta)$, where
subscripts W, S, and L refer to the wall, the solid, and the liquid
respectively. Such models qualitatively describe this system, but lose
their applicability \cite{ref1} when the size of nuclei is comparable
to the interface thickness (the nanometer range, according to
atomistic simulations \cite{ref1,ref10}). Such nanoscale nuclei are
essentially ``all interface''.  Recent investigations show \cite
{ref11} that phase field theory (PFT,\cite {ref12}) can address this
issue. Indeed, PFT can quantitatively predict the nucleation barrier
for systems (e.g, hard-sphere, Lennard-Jones, ice-water) where the
necessary input data are available. We therefore adopt this approach
to describe heterogeneous nucleation. Experimentally, the details of
the wall-fluid interaction are embedded in more directly accessible
quantities, such as the contact angle in equilibrium.  It is thus
desirable to develop a model that describes the wall in such
phenomenological terms. Along this line, interaction between 
dendritic growth and wall has recently been discussed in \cite{ref12a}, 
while Castro addressed crystal nucleation in a 
specific case of $\theta = 90^\circ$, obtained 
by prescribing "no-flux" boundary condition at the wall \cite{ref13}. 
A more general treatment is, however, required.

In this Letter, we describe how to implement phase field methods of
heterogeneous nucleation with an arbitrary contact angle. For
simplicity, we consider a single component system, whose local state
is characterized by the non-conserved phase field $\phi({\bf r})$,
defined so that 0 and 1 correspond to the bulk solid and the liquid
phases, respectively. Following previous work \cite{ref4,ref6}, we
assume that the interaction of the wall with the solidifying system is
of sufficiently short range to be characterized by a "contact free
energy" $\gamma_W(\phi)$ that depends only on the local state of
matter abutting the wall.  Accordingly, the free
energy of the system consists of a surface and a volumetric
contribution
\begin{equation}
   F_{\rm tot} = \int\limits_A dA \gamma_W(\phi)
   + \int\limits_V dV \left[  {{\epsilon^2T}\over{2}}
   (\nabla \phi)^2 + f(\phi) \right].
\end{equation}
Here $A$ is a closed surface bounding volume $V$ of the solid-liquid
system. At $A$ the system is in contact with the wall. The volumetric
term is the standard form found in thermodynamically consistent
formulations of PFT \cite{ref14}. Specifically, the local free energy
density, given the temperature $T$, is $f(\phi) = wTg(\phi) +
[1-p(\phi)] f_S(T) + p(\phi) f_L(T)$, while the "double well" and
"interpolation" functions have the forms $g(\phi) = (1/4)\phi^
2(1-\phi)^2$ and $p(\phi) = \phi^3 (10-15\phi + 6\phi^2)$. The model
parameters can be related to the solid-liquid interface free energy,
the interface thickness $\delta$ and the melting temperature $T_m$ as
$\epsilon^2 = 2^{1/2}6\gamma_{SL}\delta/T_m$ and $w
=2^{1/2}6\gamma_{SL}/(T_m\delta)$.

The critical fluctuation (nucleus) represents an extremum of the
free energy. The extremum condition $\delta F_{tot}=0$ yields
the following equations
\begin{eqnarray}
&&{{\partial\gamma_W}\over{\partial\phi}}-\epsilon^2T({\bf n} \cdot  
\nabla\phi) = 0\quad \mathrm{on\ }A;\\
&&{{\partial f}\over{\partial\phi}}-\epsilon^2T(\nabla^2\phi) = 0.
\end{eqnarray}
Here {\bf n} is the  normal vector pointing away from the wall, and 
Eq. (2)  is the boundary condition on $A$ to
Eq. (3), which in turn is the differential Euler-Lagrange (EL) equation
to be satisfied inside the volume $V$.

For simplicity, we consider first a semi-infinite system in contact
with a wall. We label the phase field at the wall $\phi_0$, and in the
far field $\phi_\infty$.  There are three possible extrema of the free
energy in this case: (i) stable solid in contact with the wall
($\phi_{\infty} = 0$; absolute minimum); (ii) metastable liquid in
contact with the wall ($\phi_{\infty}= 1$; local minimum); (iii)
unstable solid droplet  (critical fluctuation) formed in metastable
liquid at the wall ($\phi_{\infty}= 1$; saddle point).

To advance further, we must specify the contact free energy
$\gamma_W(\phi)$, this is, in general, based on the details of the
wall-fluid interaction. Thus, we now consider two illustrative choices
for our boundary conditions, and relate these choices to the
equilibrium contact angle.

{\bf Model A:} We assume that the wall does not perturb the
structure of the planar solid-liquid interface. Then $|\nabla\phi|$
can be calculated
using the one dimensional version of the integral of Eq. (3) (at $T=T_m$):
$(\epsilon^2T/2)(\nabla\phi)^2 = f(\phi) -f(\phi_{\infty}) = \Delta
f(\phi)$, and the normal component of the gradient at the wall can 
be expressed  as ${\bf n} \cdot \nabla\phi = | \nabla\phi | \cos(\theta)$.
Combining these expressions we have
\begin{equation}
{\bf n}\cdot \nabla\phi =
[\cos(\theta)/(2^{1/2}\delta)]\phi_0(1-\phi_0);\quad \mathrm{on}\ A,
\end{equation}
a condition that coincides with \cite{ref13} for $\theta = 90^\circ$.
The respective contact free energy, obtained by integrating Eq. (2),
reads as $\gamma_W(\phi)-\gamma_{WL} =
-\gamma_{SL}\cos(\theta)[2\phi^3-3\phi^2 + 1]$. Given the postulated
relationship between $\cos(\theta)$ and the interfacial free energies,
we find $\gamma(\phi)=\gamma_{WS}-\gamma_{WL}$ at the wall-solid
contact and $\gamma(\phi)=0$ at the wall-liquid contact. We adopt Eq.
(4) and the respective $\gamma_W(\phi)$ in the undercooled state.
Model A can thus be viewed as a phase field implementation of the
classical spherical cap model similar in spirit to that by Semoroz 
{\it et al.} \cite{ref12a}.

{\bf Model B:} Alternatively, we may specify $\phi_0 =
\mathrm{const.}$.  In this case only the EL equation follows from the
extremum condition. This restriction implies that $\gamma_W(\phi_0)
=\gamma_0$, independent of whether solid or liquid is in contact with
the wall. Accordingly, for planar interfaces at the melting point, the
interfacial free energies can be expressed as $\gamma_{WL} =
h(\phi_0,1) + \gamma_0, \gamma_{WS} = h(0,\phi_0) + \gamma_0$, and
$\gamma_{SL} = h(0,1)$, where
\begin{equation}
   h(\phi_1,\phi_2) = \int\limits_{\phi_1}^{\phi_2} d\phi \left[ 2 
\epsilon^2T
   \Delta f(\phi) \right]^{1/2}.
\end{equation}
After some algebra, we find
$\cos(\theta)= (\gamma_{WL}-\gamma_{WS}) / \gamma_{LS}
=  1 -  6 \phi_0^2 + 4 \phi_0^3$.  The condition that  $\phi_0 =
\mathrm{const.}$  sets the degree of ordering (or
disordering) of the substance abutting the wall. In other words we see
{\it liquid ordering} near the wall if liquid phase is assumed in the  
far field,
while a {\it disordered solid} forms at the wall if solid is assumed
in the far field. Inspection of the integral EL equation indicates
that the metastable liquid solution exists so far as $\Delta f 
(\phi_0) > 0$,
i.e., below a critical undercooling $\Delta T_c$ determined by the
condition $\Delta f(\phi_0) = 0$. (As $\phi_0$ approaches the
solid state, $\Delta T_c$ converges
to 0.) At lower $T$, there is no
time-independent solution of this type, instead a propagating  
solidification
front emerges that is described by the usual equation of motion
for the phase field \cite{ref14}. This 
model does not converge to \cite{ref13} at $\theta = 90^\circ$.

Our two choices of the boundary condition at the wall correspond to
two distinct physical situations: (a) The "classical" case, when
liquid ordering is negligible at the wall, and (b) a "non-classical"
case, where there is an imposed order at the wall.  This structure is
of a specific nature, as it corresponds the particular, chosen level
of ordering as one traverses solid-liquid interface.  As such, this
order is "compatible" with the appearing crystal structure, and will
lower the nucleation barrier to the formation of solid.  While it is
typical for liquids to order at an abutting wall [7(c), 10(b), 10(d)],
such ordering may not be compatible with the structure to which the
liquid crystallizes \cite{ref3}, and a more detailed model would be
required. Based on these observations, we expect that our combined
analyses of Models A and B will elucidate many of the essential
behaviors of physical systems.  In what follows, we evaluate the
properties of heterogeneous nuclei in these two limiting cases,
 and present illustrative model simulations for pure Ni \cite{ref15}.

The EL equation for the composite system nucleus plus undercooled
liquid has been solved by the finite element method.  The initial
condition has been created by placing the classical sharp interface
nucleus into the simulation window after broadening its interface by a
$\tanh$ function of appropriate width parameter.  The simulation box
had the size of 30 nm $\times$ 20 nm. The equation of motion for
dynamic evolution simulations was solved in a dimensionless form using
the finite difference method and parallel computing on a PC cluster of
120 nodes.  The spatial step was $\Delta x = 0.2$ nm, while noise (as
described in \cite{ref13}), has been added to the governing equation.

\begin{figure}
\includegraphics[width=8cm]{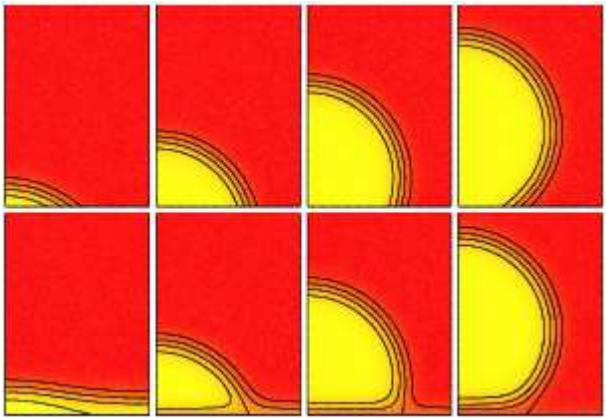}
\caption{(Color online). Structure of heterogeneous nuclei in 2D
vs. equilibrium contact angle at $\Delta T$ = 35 K in Model A 
(upper row) and B (lower row). From left to right $\theta$ = 
37.6$^\circ$, 72.8$^\circ$, 107.2$^\circ$, and 142.3$^\circ$ 
($\phi_0$ = 0.2, 0.4, 0.6, and 0.8). The contour lines stand for 
$\phi$ = 0.2, 0.4, 0.6, and 0.8. Horizontal size is 10 nm. Coloring: 
red -- liquid; yellow -- solid. For symmetry reasons, only the right 
half of the nuclei is shown.}
\end{figure}

\begin{figure}[t]
\includegraphics[width=8cm]{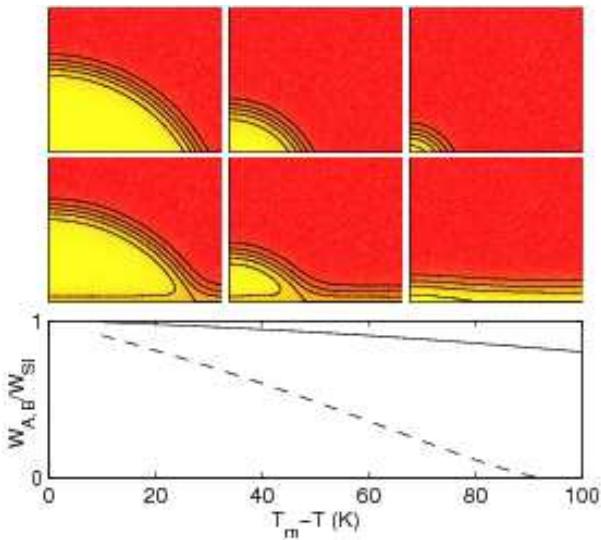}
\caption{(Color online). Structure of nuclei in 2D at three 
undercoolings ($\Delta T$ = 20 K, 40 K, and 90 K) in 
Models A (upper row) and B (central row),
at an equilibrium contact angle of $\theta=61.2^\circ$ corresponding 
to $\phi_0=1/3$  and $\Delta T_c =$ 92.0 K. The contour lines 
indicate $\phi=1/6, 2/6, 3/6, 4/6$, and 5/6, respectively. 
Horizontal size is 12 nm. Coloring as for Fig. 1. The free energy 
of non-classical nuclei $(W_{A,B})$ normalized by the sharp interface 
prediction $(W_{SI})$
is also shown as a function of undercooling (bottom panel): 
Model A -- solid; Model B -- dashed. 
}
\end{figure}

Critical fluctuations (nuclei) computed at undercooling $\Delta T =$
35 K as a function of the equilibrium contact angle $\theta$ 
are presented in Fig. 1. 
For comparison, Fig. 2 shows the nuclei calculated for a
contact angle of $\theta$ = 61.2$^\circ$ ($\phi_0 = 1/3$) as a 
function of $\Delta T$. While in both models the size of
heterogeneous nuclei becomes comparable to the interface thickness
with increasing undercooling, in Model B it happens at a far smaller
undercooling. It is remarkable that while the contact angle
is approximately constant in Model A, in Model B it  varies drastically with
undercooling in Model B, and tends to 0$^\circ$ complete wetting at
the critical undercooling. The free energy of heterogeneous nuclei in
Model A and B are also shown in Fig. 2.  For Model A, it has been
calculated by integrating the free energy density difference relative
to the bulk undercooling liquid and adding the contribution from the
wall. For Model B, the integrated free energy of the wall-liquid
system has been subtracted from the free energy of the
wall-nucleus-liquid system. It is found that the free energies of
nuclei from Models A and B fall close to the values from the sharp
interface spherical cap model (2D) for small undercoolings where the
nuclei are large relative to the interface thickness, while lower
values are obtained at larger undercoolings. In Model B, the
nucleation barrier disappears at $\Delta T_c$. Atomistic simulations
could test the existence of such a $\Delta T_c$.

\begin{figure}[b]
\includegraphics[width=8cm]{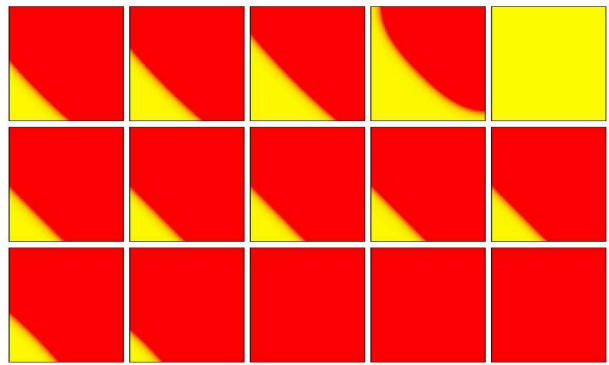}
\caption{(Color online). Wetting of rectangular corner by the 
crystalline phase at $T = T_m$ in Model A.  
From up to down the contact angle is 40$^\circ$, 45$^\circ$ (critical), 
and 50$^\circ$, respectively. Time increases to the right. Snapshots 
were taken at 5000, 20,000, 40,000, 60,000 and 80,000 dimensionless 
time steps. (Coloring as for Fig. 1. 100$\times$100 grid.)
Similar results were obtained for Model B.}
\end{figure}

\begin{figure*}[t]
\includegraphics[width=17cm]{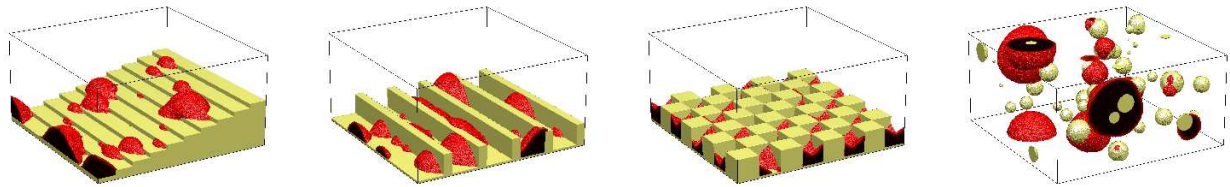}
\caption{(Color online). Phase field simulation of heterogeneous 
nucleation of Ni crystals  
on complex surfaces in 3D using Model A. From left to right: 
on stairs, in rectangular grooves, on a checkerboard-modulated 
surface, and on floating nanoparticles.
[$\Delta T =$ 300 K. A $400 \times 400 \times 200$ grid 
corresponding to 80 nm $\times$ 80 nm $\times$ 40 nm is used. Coloring:
 solid-liquid interface (defined by $\phi = 0.5)$ -- red; bulk 
crystal -- black; wall -- yellow.]}
\end{figure*}

Next, we demonstrate that these models describe wetting as it
is usually understood.  For a corner of angle $\alpha$, there exists a
critical contact angle $\theta_c = \pi/2 - \alpha/2$ below which there
is no nucleation barrier.  For a right angle, or square corner, $\theta_c$
= 45$^\circ$.  At the melting point we examine three cases in a square
corner: (a) We select a wetting angle of $\theta=\theta_c$.  An
isosceles triangle crystal formed in this corner realizes a planar
solid-liquid interface. This interface is stable, since the capillary
pressure vanishes for a planar interface. (b) For $\theta < \theta_c$,
the solid-liquid interface is concave, and resulting in a negative
capillary pressure that increases the melting point (Gibbs-Thomson
effect), so a crystal triangle posited in the corner must grow. (c) For
$\theta > \theta_c$, a convex solid-liquid interface develops and a
crystal triangle placed in the corner melts. Simulations performed for
rectangular corner display the expected behavior for both models
(Fig. 3).

Illustrative simulations demonstrating the power of our
approach, are presented in Fig. 4.  Using Model A,  with $\theta =
60^\circ$,  heterogeneous nucleation is modeled on stairs, in rectangular 
grooves, on a checkerboard-modulated surface, and on
floating nanoparticles. Simulations of this kind are expected 
to find applications in nanopatterning studies \cite{ref0}.

We have developed a phase field methodology to
describe heterogeneous crystal nucleation in undercooled liquids at
walls characterized by arbitrary contact angles. Two limiting cases
have been addressed: (Model A) Nucleation at surfaces where liquid 
ordering at the wall is negligible and (Model B) where the wall-liquid 
interaction induces partial crystalline order in the liquid (Model B). 
Using the prescriptions described above, many other boundary conditions 
can be explored. Note that this approach can be applied to any system displaying
a first order phase transition (such as vapor-liquid).  Also, this
approach can be directly extended to existing phase field models of
alloy and anisotropic polycrystalline systems characterized by further
fields. The present study thus opens up new ways for modeling
heterogeneous nucleation in a broad variety of systems.

The authors acknowledge helpful discussions with
J. W. Cahn and G. B. McFadden. This work has been supported 
by contracts OTKA-T-037323, OTKA-K-62588,
ESA PECS No. 98005, and by the EU FP6 Project IMPRESS
under Contract No. NMP3-CT-2004-500635. T. P. acknowledges support
by the Bolyai J\'anos Scholarship.


\begin{thebibliography}{99}
\bibitem{ref15}
  Properties of pure Ni we used: $T_m =$ 1728 K, 
  heat of fusion $L = 2.666\times 10^9$ J/m$^3$,  $\gamma_{SL} =$ 0.364 J/m$^2$, 
  free energy difference $f_L - f_S =  L (T_m - T)/T_m$. The $10\%-20\%$ interface 
  thickness was assumed to be $d =$ 2 nm.
\end{thebibliography}
\end{document}